\documentclass[pra,twocolumn,showpacs,preprintnumbers,amsmath,amssymb]{revtex4}

\usepackage{graphicx}
\usepackage{dcolumn}
\usepackage{bm}

\begin{document}

\title{Performance Analysis for Brandt's Conclusive Entangling Probe}

\author{Jeffrey H. Shapiro}
\email[Electronic address: ]{jhs@mit.edu}
\affiliation{Massachusetts Institute of Technology, Research Laboratory of Electronics, Cambridge, Massachusetts 02139 USA}

\date{\today}

\begin{abstract}
The Fuchs-Peres-Brandt (FPB) probe realizes the most powerful individual attack on Bennett-Brassard 1984  quantum key distribution by means of a single controlled-NOT gate in which Alice's transmitted qubit becomes the control-qubit input, Bob's received qubit is the control-qubit output, and Eve supplies the target-qubit input and measures the target-qubit output.  The FPB probe uses the minimum-error-probability projective measurement for discriminating between the target-qubit output states that are perfectly correlated with Bob's sifted bit value when that bit is correctly received.  This paper analyzes a recently proposed modification of the FPB attack in which Eve's projective measurement is replaced by a probability operator-valued measurement chosen to unambiguously discriminate between the same two target-qubit output states.
\end{abstract}

\maketitle

\section{Introduction}   
In an individual attack on single-photon, polarization-based Bennett-Brassard 1984 quantum key distribution (BB84 QKD), Eve probes Alice's photons one at a time.   Fuchs and Peres \cite{FP} described the most  general way in which an individual attack could be performed.  Eve supplies a probe photon and lets it interact with Alice's photon in a unitary manner.   Eve then sends Alice's photon  to Bob, and performs a probability operator-valued measurement (POVM) on the probe photon she has retained.  Slutsky {\em et al.} \cite{Slutsky} demonstrated that the Fuchs-Peres construct---with the appropriate choice of probe state, interaction, and measurement---affords Eve the maximum amount of R\'{e}nyi information about the error-free sifted bits that Bob receives for a given level of disturbance, i.e., for a given probability that a sifted bit will be received in error.  Brandt \cite{FPB} extended the Slutsky {\em et al.} treatment by showing that the optimal probe could be realized with a single CNOT gate; see Fig.~1 for an abstract diagram of the resulting Fuchs-Peres-Brandt (FPB) probe.  Shapiro and Wong \cite{PRA} then showed how a complete physical simulation of the FPB probe could be accomplished using single-photon two-qubit (SPTQ) quantum logic \cite{SPTQ}, and how this arrangement could be transformed into a true deterministic realization of the FPB attack once polarization-insensitive quantum nondemolition photon detection is developed.  
\begin{figure}[h]
\includegraphics[width = 3.4in]{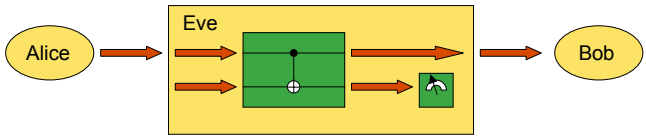}
\caption{Block diagram of the Fuchs-Peres-Brandt probe for attacking BB84 QKD.}
\end{figure}

Consider ideal conditions, in which Alice transmits a single photon per bit interval, there is no propagation loss and no extraneous (background) light collection, and both Eve and Bob have unity quantum efficiency photodetectors with no dark counts.  These conditions imply there will not be any errors on sifted bits---photons for which Alice and Bob have chosen the same polarization basis---in the absence of eavesdropping.  As discussed in \cite{FPB},\cite{PRA},\cite{JMO}, when Eve's intrusion causes Bob to suffer an error probability $P_E$ on his sifted bits, Eve derives a R\'{e}nyi information equal to 
\begin{equation}
I_R = \log_2\!\left(1 + \frac{4P_E(1-2P_E)}{(1-P_E)^2}\right),
\label{Renyi}
\end{equation}
about each of Bob's error-free sifted bits.  Thus, Eve's R\'{e}nyi information climbs from 0 to 1 bit per error-free sifted bit when $P_E$ increases from 0 to 1/3, as shown in Fig.~2.  
\begin{figure}[h]
\includegraphics[width = 2in]{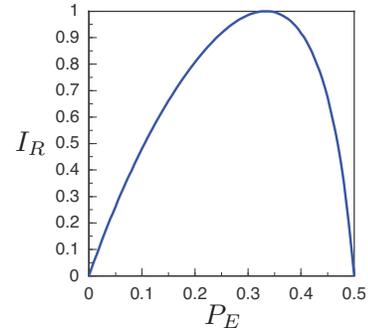}
\caption{Eve's R\'{e}nyi information---obtained from the Fuchs-Peres-Brandt probe---about Bob's error-free sifted bits versus the error probability that her eavesdropping creates.}
\end{figure}

The preceding performance tradeoff between the error probability that Eve creates by her eavesdropping versus the R\'{e}nyi entropy she accrues derives from the effect of her CNOT gate on Alice's photon and the projective measurement Eve makes on the target-qubit output from that CNOT gate.  The projective measurement is chosen for minimum-error-probability discrimination between the two target-qubit output states that are perfectly correlated with Bob's sifted bit value when that bit is correctly received.  For $P_E< 1/3$, these target-qubit output states are not orthogonal, hence Eve's R\'{e}nyi information is less than perfect in this regime, because non-orthogonal states cannot be distinguished without error.  For $P_E >0$, Eve's CNOT gate disturbs Alice's photon---the control qubit in Fig.~1---en route to Bob, so that her eavesdropping inevitably injects errors into the otherwise ideal connection between Alice and Bob.  

Recently, Brandt has introduced the notion of a conclusive entangling probe \cite{QIP}--\cite{arx2}.  In essence, this is the Fig.~1 configuration with Eve's projective measurement replaced by a POVM chosen to unambiguously discriminate \cite{Ekert} between the same two target-qubit output states considered previously.    Brandt's performance analysis for the conclusive entangling probe neglects the effect of the the third possible output state for Eve's target qubit, viz., the state that heralds Bob's receiving a sifted bit in error.  Thus the purpose of the present paper is to evaluate the performance of Brandt's conclusive entangling probe when that third state is included in the analysis, as it must be.  Our work will show that this performance is dramatically different than what Brandt derived by inappropriately ignoring the third target-qubit output state.  We begin, in Sec.~II, by reviewing the projective-measurement FPB probe, using the notation from \cite{PRA}.  We continue, in Sec.~III, with our analysis of Brandt's conclusive entangling probe, and conclude, in Sec.~IV, by discussing the implications of our results.  

\section{The Fuchs-Peres-Brandt Probe}
Under the ideal operating conditions that we have presumed, the FPB attack on single-photon, polarization-based BB84 QKD proceeds as follows.  In each bit interval Alice transmits, at random, a single photon in one of the four BB84 polarization states, i.e., $|H\rangle, |V\rangle, |$$+$$45^\circ\rangle$ or $|$$-$$45^\circ\rangle$.  Eve uses this photon as the control-qubit input to a CNOT gate whose computational basis---relative to the BB84 polarization states---is shown in Fig.~2, namely 
\begin{eqnarray}
|0\rangle &\equiv& \cos(\pi/8)|H\rangle + \sin(\pi/8)|V\rangle \\ 
|1\rangle &\equiv& -\sin(\pi/8)|H\rangle + \cos(\pi/8)|V\rangle,
\end{eqnarray} 
in terms of the $H/V$ basis.  Eve supplies her own probe photon, as the target-qubit input to this CNOT gate, in the state
\begin{equation}
|T_{\rm in}\rangle \equiv C|+\rangle + S|-\rangle,
\label{probeinput}
\end{equation}
where $C = \sqrt{1-2P_E}$, $S = \sqrt{2P_E}$, $|\pm\rangle = (|0\rangle \pm |1\rangle)/\sqrt{2}$, and $0\le P_E\le 1/2$ will turn out to be the error probability that Eve's probe creates on Bob's sifted bits.  So, as $P_E$ increases from 0 to 1/2, $|T_{\rm in}\rangle$ goes from $|+\rangle$ to $|-\rangle$.  The (unnormalized) output states that may occur for this target qubit are
\begin{eqnarray}
|T_\pm\rangle &\equiv& C|+\rangle \pm \frac{S}{\sqrt{2}}|-\rangle \\ 
|T_E\rangle &\equiv& \frac{S}{\sqrt{2}}|-\rangle.
\end{eqnarray}
\begin{figure}
\vspace*{.1in}
\includegraphics[width = 2.4in]{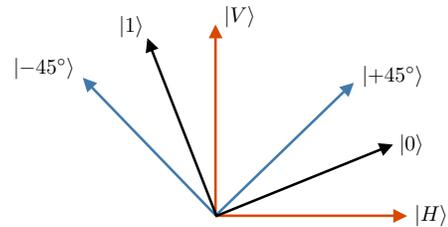}
\caption{Computational basis for Eve's CNOT gate referenced to the BB84 polarization states.}
\end{figure}

When Alice uses the $H/V$ basis for her photon transmission, Eve's CNOT gate effects the following transformation,
\begin{eqnarray}
|H\rangle|T_{\rm in}\rangle &\longrightarrow&
|H\rangle|T_-\rangle + |V\rangle|T_E\rangle \label{Hin_out} 
\\ 
|V\rangle|T_{\rm in}\rangle &\longrightarrow&
|V\rangle|T_+\rangle +|H\rangle|T_E\rangle,\label{Vin_out} 
\end{eqnarray}
where the kets on the left-hand side denote the Alice\,$\otimes$\,Eve state of the control and target qubits at the CNOT's input and the kets on the right-hand side denote the Bob\,$\otimes$\,Eve state of the control and target qubits at the CNOT's output.  
Similarly, when Alice uses the $\pm 45^\circ$ basis, Eve's CNOT gate has the following behavior,
\begin{eqnarray}
|\mbox{$+$}45^\circ\rangle|T_{\rm in}\rangle &\longrightarrow&
|\mbox{$+$}45^\circ\rangle|T_+\rangle + |\mbox{$-$}45^\circ\rangle|T_E\rangle 
\label{plus_in_out}\\ 
|\mbox{$-$}45^\circ\rangle|T_{\rm in}\rangle &\longrightarrow&
|\mbox{$-$}45^\circ\rangle|T_-\rangle +|\mbox{$+$}45^\circ\rangle|T_E\rangle.
\label{minus_in_out}
\end{eqnarray}
Suppose that Bob measures in the basis that Alice has employed \em and\/\rm\ his outcome matches what Alice sent.  Then Eve can learn their shared  bit value, once Bob discloses his measurement basis, by distinguishing between the $|T_+\rangle$ and $|T_-\rangle$ output states for her target qubit.  Of course, this knowledge comes at a cost:  Eve has caused an error event whenever Alice and Bob choose a common basis and her target qubit's output state is $|T_E\rangle$.  To maximize the information she derives from this intrusion, Eve applies the minimum error probability receiver for distinguishing between
the single-photon polarization states $|T_+\rangle$ and $|T_-\rangle$.  This is a projective measurement onto the polarization basis $\{|d_+\rangle,|d_-\rangle\}$, shown (for $0< P_E <1/3$) in Fig.~3 and given by
\begin{eqnarray}
|d_+\rangle &=& \frac{|+\rangle + |-\rangle}{\sqrt{2}} = |0\rangle \\ 
|d_-\rangle &=& \frac{|+\rangle - |-\rangle}{\sqrt{2}} = |1\rangle. 
\end{eqnarray}
\begin{figure}[h]
\includegraphics[width = 2.2in]{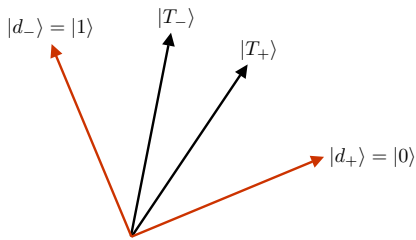}
\caption{Measurement basis for Eve's minimum-error-probability discrimination between $|T_+\rangle$ and $|T_-\rangle$; figure assumes $0<P_E<1/3$.}
\end{figure}
Straightforward calculations \cite{FPB},\cite{PRA},\cite{JMO} now show that $P_E$ is the error probability that Eve's presence creates on Bob's sifted bits, and $I_R$ from Eq.~(\ref{Renyi}) is the R\'{e}nyi information that Eve derives about Bob's error-free sifted bits.

\section{Conclusive Entangling Probe}
Suppose, as in \cite{QIP}--\cite{arx2}, that Eve replaces the projective measurement $\{|d_+\rangle, |d_-\rangle\}$ in her FPB probe with the POVM
\begin{eqnarray}
\hat{\Pi}_+ &=&  \frac{|T_-^\perp\rangle\langle T_-^\perp|}{2C^2} \\[.05in] 
\hat{\Pi}_- &=& \frac{|T_+^\perp\rangle\langle T_+^\perp|}{2C^2}\\[.05in] 
\hat{\Pi}_{\rm inc} &=&  \frac{C^2 - S^2/2}{C^2}\,|+\rangle\langle +|  \\[.05in]
&=& \hat{I} - \hat{\Pi}_+ -\hat{\Pi}_-,
\end{eqnarray}
where $\hat{I}$ is the identity operator on the target qubit's state space, the (unnormalized) states
\begin{equation}
|T_\pm^\perp\rangle \equiv -\frac{S}{\sqrt{2}}|+\rangle \pm C|-\rangle
\end{equation} 
are orthogonal to $|T_\pm\rangle$, respectively, see Fig.~5, and $P_E$  has been restricted to the interval $0\le P_E\le 1/3$.  Although our analysis of the conclusive entangling probe can be modified to cover an attack with $1/3< P_E\le 1/2$, we shall not do so.  Our reason for eschewing the high-$P_E$ regime is that an intelligent Eve would never mount a projective-measurement FPB attack with $P_E > 1/3$ because, as shown in Fig.~2, her R\'{e}nyi information decreases---from a 1 bit per error-free sifted bit to 0---as $P_E$ increases from 1/3 to 1/2.  
\begin{figure}[h]
\vspace*{.2in}
\includegraphics[width = 2.2in]{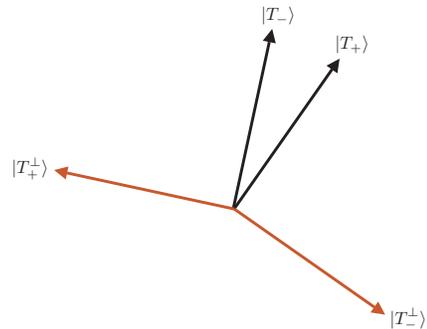}
\caption{Measurement basis for Eve's conclusive-probe POVM discrimination between $|T_+\rangle$ and $|T_-\rangle$; figure assumes $0< P_E <1/3$.}
\end{figure}

Let the outcomes Eve associates with the POVM elements $\hat{\Pi}_+,\hat{\Pi}_-$, and $\hat{\Pi}_{\rm inc}$ be denoted $T_+, T_-$, and $T_{\rm inc}$.  They represent, respectively, Eve's thinking that her target qubit's output state was $|T_+\rangle$, $|T_-\rangle$, or that her  measurement was inconclusive.  When Alice and Bob have both used the $H/V$ basis, Eve will decode her $T_+$ outcome as $V$, and her $T_-$ outcome as $H$, after that basis choice is revealed.   Likewise, when Alice and Bob have both used the $\pm 45^\circ$ basis, Eve will decode her $T_+$ outcome as $+$45$^\circ$ and her $T_-$ outcome as $-$45$^\circ$ after basis revelation.  It is now a simple matter to evaluate the joint Alice ($A$), Bob ($B$), and Eve ($E$) probabilities that we will need to understand the performance of Brandt's conclusive entangling probe.  Because only bits for which Alice and Bob have chosen the same polarization basis need be considered, all of the probabilities we shall evaluate and employ below will be conditioned on there being an Alice/Bob sift event.  However, for notational simplicity, we shall not carry that conditioning along explicitly.  Also, for the sake of brevity, we will only present the calculation details for the case in which Alice sends an $H$-polarized photon.  The complete set of $ABE$ sifted-bit joint probabilities appears  in Table~I.

When there is a sift event in which Alice has sent an $H$-polarized photon, the joint state of Bob's qubit and Eve's, found from Eq.~(\ref{Hin_out}), is
\begin{equation}
|\psi_H\rangle \equiv |H\rangle|T_-\rangle + |V\rangle|T_E\rangle, 
\end{equation}
and Bob measures his qubit with the projectors
\begin{eqnarray}
\hat{\Pi}_H &=& |H\rangle\langle H| \\ 
\hat{\Pi}_V &=& |V\rangle\langle V|.
\end{eqnarray}
Thus, the joint conditional probability that Bob and Eve will both measure $H$ is found as follows:
\begin{eqnarray}
\lefteqn{\Pr(\,B = H, E = H \mid A = H\,) = } \nonumber \\ 
& &  \langle \psi_H |(\hat{\Pi}_H\otimes\hat{\Pi}_-)|\psi_H\rangle
= \frac{|\langle T_-|T_+^\perp\rangle|^2}{2C^2} = 2P_E.
\end{eqnarray}
Proceeding in the same way we can obtain the other joint conditional probabilities for the $A=H$ case:
\begin{eqnarray}
\lefteqn{\Pr(\,B = V, E = H \mid A = H\,) = } \nonumber \\ 
& &  \langle \psi_H |(\hat{\Pi}_V\otimes\hat{\Pi}_-)|\psi_H\rangle
= \frac{|\langle T_E|T_+^\perp\rangle|^2}{2C^2} = P_E/2 \\[.05in] 
\lefteqn{\Pr(\,B = H, E = V \mid A = H\,) = } \nonumber \\ 
& &  \langle \psi_H |(\hat{\Pi}_H\otimes\hat{\Pi}_+)|\psi_H\rangle
= \frac{|\langle T_-|T_-^\perp\rangle|^2}{2C^2} = 0 \\[.05in] 
\lefteqn{\Pr(\,B = V, E = V \mid A = H\,) = } \nonumber \\ 
& &  \langle \psi_H |(\hat{\Pi}_V\otimes\hat{\Pi}_+)|\psi_H\rangle
= \frac{|\langle T_E|T_-^\perp\rangle|^2}{2C^2} = P_E/2\\[.05in] 
\lefteqn{\Pr(\,B = H, E = {\rm inc} \mid A = H\,) = } \nonumber \\ 
&=& \langle \psi_H|(\hat{\Pi}_H\otimes\hat{\Pi}_{\rm inc})|\psi_H\rangle = 
\frac{C^2 - S^2/2}{C^2}\,|\langle T_-|+\rangle|^2 \nonumber \\[.05in] 
&=& 1 -3 P_E\\[.05in]
\lefteqn{\Pr(\,B = V, E = {\rm inc} \mid A = H\,) = } \nonumber \\ 
&=& \langle\psi_H|(\hat{\Pi}_V\otimes\hat{\Pi}_{\rm inc})|\psi_H\rangle 
= \frac{C^2 - S^2/2}{C^2}\,|\langle T_E|+\rangle|^2  \nonumber\\[.05in] 
&=& 0.
\end{eqnarray}
Conditioned on there being a sift event, the $ABE$ joint probabilities in which $A=H$ are gotten by multiplying the preceding conditional probabilities by $\Pr(A = H) = 1/4$.  Similar calculations yield the full set of $ABE$ joint probabilities given in Table~I.  Because this table includes all $ABE$ possibilities that are consistent with the occurrence of an Alice/Bob sift event, its probabilities---which are conditioned on there being such a sift event---sum to one.
\begin{table}[h]
\begin{center}
\begin{tabular}{|c|c|c||c|}\hline
Alice & Bob & Eve & Probability\\ \hline \hline
$H$ & $H$ & $H$ & $P_E/2$\\ \hline
$H$ & $V$ & $H$ & $P_E/8$ \\ \hline
$H$ & $H$ & $V$ & 0 \\ \hline
$H$ & $H$ & inc & $(1-3P_E)/4$ \\ \hline
$H$ & $V$ & inc & 0 \\ \hline
$H$ & $V$ & $V$ & $P_E/8$ \\ \hline
$V$ & $V$ & $V$ & $P_E/2$ \\ \hline
$V$ & $H$ & $V$ & $P_E/8$ \\ \hline
$V$ & $V$ & $H$ & 0 \\ \hline
$V$ & $H$ & $H$ & $P_E/8$ \\ \hline
$V$ & $V$ & inc & $(1-3P_E)/4$ \\ \hline
$V$ & $H$ & inc & 0 \\ \hline
$+$45$^\circ$ & $+$45$^\circ$ & $+$45$^\circ$ & $P_E/2$ \\ \hline
$+$45$^\circ$ & $-$45$^\circ$ & $+$45$^\circ$ & $P_E/8$ \\ \hline
$+$45$^\circ$ & $+$45$^\circ$ & $-$45$^\circ$ & 0 \\ \hline
$+$45$^\circ$ & $-$45$^\circ$ & $-$45$^\circ$ & $P_E/8$ \\ \hline
$+$45$^\circ$ & $+$45$^\circ$ & inc & $(1-3P_E)/4$ \\ \hline
$+$45$^\circ$ & $-$45$^\circ$ & inc & 0 \\ \hline
$-$45$^\circ$ & $-$45$^\circ$ & $-$45$^\circ$ & $P_E/2$ \\ \hline
$-$45$^\circ$ & $+$45$^\circ$ & $-$45$^\circ$ & $P_E/8$ \\ \hline
$-$45$^\circ$ & $-$45$^\circ$ & $+$45$^\circ$ & 0 \\ \hline
$-$45$^\circ$ & $+$45$^\circ$ & $+$45$^\circ$ & $P_E/8$ \\ \hline
$-$45$^\circ$ & $-$45$^\circ$ & inc & $(1-3P_E)/4$ \\ \hline
$-$45$^\circ$ & $+$45$^\circ$ & inc & 0 \\ \hline
\end{tabular}
\end{center}
\caption{Joint probabilities for Alice, Bob, and Eve conditioned on there being an Alice/Bob sift event and assuming $0\le P_E\le 1/3$.} 
\end{table}

The probabilities listed in Table~I have three interesting characteristics.  Eve is using the POVM for unambiguous discrimination between $|T_+\rangle$ and $|T_-\rangle$.  Thus, whenever Eve's measurement outcome is $T_+$ or $T_-$ and Bob's sifted bit value agrees with Alice's, Eve will \em not\/\rm\ make an error, e.g., $\Pr(A=H, B=H, E=V) = 0$, in keeping with the notion of unambiguous detection. It might also be expected \cite{QIP}--\cite{arx2} that when Eve's measurement outcome is $T_+$ or $T_-$---i.e., when it is \em not\/\rm\ $T_{\rm inc}$ so that she thinks her measurement is conclusive---that Bob never suffers a sifted-bit error.  Table~I shows that such is \em not\/\rm\  the case, e.g., $\Pr(A = H, B = V, E = H) = P_E/8$ is non-zero for $0<P_E\le 1/3$.  This deviation from unambiguous detection occurs because Eve's target qubit may be in one of \em three\/\rm\ different states:  $|T_+\rangle$, $|T_-\rangle$, or $|T_E\rangle$.  The presence of $|T_E\rangle$, which is not orthogonal to $|T_+^\perp\rangle$ or $|T_-^\perp\rangle$ for $0 < P_E \le /1/3$, is directly responsible for Bob's sometimes receiving sifted bits that differ from Alice's.  On the other hand, when Eve's measurement outcome is $T_{\rm inc}$, Alice and Bob have no errors on their sifted bits.  

The quantities we need to complete our performance analysis of Brandt's conclusive entangling probe can all be obtained from Table~I.    Let ${\cal{A}} = \{0,1\}$, ${\cal{B}} = \{0,1\}$ and ${\cal{E}} = \{0,1, {\rm inc}\}$ denote the ensembles of possible bit values that Alice sends, and Bob and Eve receive, where we assume the following $H/V$ and $\pm 45^\circ$ encodings:
\begin{eqnarray}
&&0\rightarrow H, 1\rightarrow V,\quad \mbox{for $H/V$ basis}\\
&&0\rightarrow +45^\circ, 1\rightarrow -45^\circ,\quad\mbox{for $\pm45^\circ$ basis}.
\end{eqnarray}
The joint $abe$ probabilities, conditioned on there being a sift event, are listed in Table~II; they were obtained by adding the appropriate Table~I probabilities in accordance with the preceding $H/V$ and $\pm 45^\circ$ encodings.

We want to find:  $\Pr(b\neq a)$, the Alice/Bob error probability; $\Pr(e = {\rm inc})$, the probability that Eve regards her measurement as inconclusive; and $I_R$, the R\'{e}nyi entropy Eve derives about Bob's error-free sift events.  From Table~II we immediately find that 
\begin{eqnarray}
\Pr(b\neq a) &=& P_E
\label{errprob}\\
\Pr(e = {\rm inc}) &=& 1- 3P_E.
\label{incprob}
\end{eqnarray}
The R\'{e}nyi information (in bits) that Eve learns about each Alice/Bob error-free sift event is 
\begin{eqnarray}
I_R &\equiv&
-\log_2\!\left(\sum_b[\Pr(\,b\mid b=a\,)]^2\right) \nonumber \\
&+&\sum_e\Pr(\,e\mid b=a\,) \nonumber \\ 
&\times& \log_2\!\left(\sum_b [\Pr(\,b\mid e, b=a\,)]^2\right).
\end{eqnarray}
From Table~II we find that
\begin{equation}
\Pr(\,b = j \mid b = a\,) = 1/2,\quad\mbox{for $j=0,1$}
\end{equation}
\begin{equation}
\Pr(\,e = k \mid b = a\,) = \left\{\begin{array}{ll}
\frac{\displaystyle P_E}{\displaystyle 1-P_E}, & \mbox{for $k = 0,1$}\\[.15in]
\frac{\displaystyle 1-3P_E}{\displaystyle 1-P_E}, & \mbox{for $k = {\rm inc}$}\end{array}\right.
\end{equation} 
\begin{equation}
\Pr(\,b = j\mid e = j, b=a\,) = 1,\quad\mbox{for $j=0,1$}
\end{equation}
and
\begin{equation}
\Pr(\,b = j \mid e = {\rm inc}, b = a\,) = 1/2,\quad\mbox{for $j = 0,1$,}
\end{equation}
whence
\begin{equation}
I_R = \frac{2P_E}{1-P_E}.
\label{renyi1}
\end{equation}
\begin{table}
\begin{center}
\begin{tabular}{|c|c|c||c|}\hline
$a$ & $b$ & $e$ & Probability\\ \hline \hline
$0$ & $0$ & $0$ & $P_E$\\ \hline
$0$ & $0$ & $1$ & $0$ \\ \hline
$0$ & $0$ & inc & $(1-3P_E)/2$ \\ \hline
$0$ & $1$ & $0$ & $P_E/4$ \\ \hline
$0$ & $1$ & $1$ & $P_E/4$ \\ \hline
$0$ & $1$ & inc & $0$ \\ \hline
$1$ & $0$ & $0$ & $P_E/4$ \\ \hline
$1$ & $0$ & $1$ & $P_E/4$ \\ \hline
$1$ & $0$ & inc & $0$ \\ \hline
$1$ & $1$ & $0$ & $0$ \\ \hline
$1$ & $1$ & $1$ & $P_E$ \\ \hline
$1$ & $1$ & inc & $(1-3P_E)/2$ \\ \hline
\end{tabular}
\end{center}
\caption{Joint bit probabilities for Alice, Bob, and Eve conditioned on there being an Alice/Bob sift event and assuming $0\le P_E\le 1/3$.} 
\end{table}

\section{Discussion}
We are now ready to evaluate the merits of Brandt's conclusive entangling probe.  For our ideal conditions, in which there is no propagation loss, no background light, and Eve and Bob have unity quantum efficiency photodetectors with no dark counts, Eqs.~(\ref{errprob}), (\ref{incprob}) and (\ref{renyi1}) characterize the performance of this system when Eve routes all of Alice's photons to Bob.  Equation~({\ref{errprob}) shows that Eve's actions have caused Alice and Bob to have an error probability of $P_E$ on their sifted bits, i.e., the same result as obtained for the FPB probe that uses the minimum-error-probability projective measurement.  This comes as no surprise:  the Alice/Bob error probability is caused by the CNOT gate, \em not\/\rm\ by the measurement that Eve makes on her output of the CNOT gate.  Equation~(\ref{incprob}) shows that Eve's probability of thinking her measurement is inconclusive can be tuned to be anywhere between 0 to 1 by appropriate choice of $P_E$ in the range $0\le P_E\le 1/3$.  When $P_E = 0$, Eve's $\Pr({\rm inc}) = 1$, because $|T_+\rangle = |T_-\rangle$ in this case.  When $P_E = 1/3$, Eve's $\Pr({\rm inc}) = 0$, because $\langle T_+|T_-\rangle = 0$ in this case.  Of greater interest is the behavior of Eve's R\'{e}nyi information when she uses the conclusive entangling probe.  To compare the information Eve receives from the FPB probe with what she gets from the Brandt's conclusive entangling probe we have plotted the R\'{e}nyi informations (in bits) from Eqs.~(\ref{Renyi}) and (\ref{renyi1}) in Fig.~6 for $0\le P_E \le 1/3$.  We see that both increase from 0 to 1 bit per error-free sifted bit as $P_E$ increases from 0 to 1/3.  Both probes yield zero R\'{e}nyi information when $P_E = 0$, because $|T_+\rangle = |T_-\rangle$ and $|T_E\rangle = 0$ in this case, i.e., Eve's target-qubit output state is \em not\/\rm\ entangled with Bob's qubit.  Both probes yield 1 bit of R\'{e}nyi information per error-free sifted bit when $P_E = 0$, because $\langle T_+|T_-\rangle = 0$ in this case.  However, for $0< P_E <1/3$, the R\'{e}nyi information obtained from the FPB probe (with its minimum-error-probability projective measurement) exceeds that found by using the conclusive entangling probe (with its POVM measurement).  
\begin{figure}
\vspace*{.2in}
\includegraphics[width = 2in]{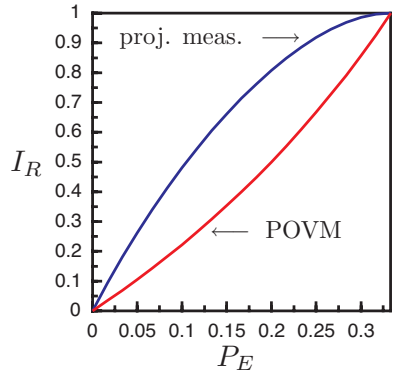}
\caption{Eve's R\'{e}nyi information about Bob's error-free sifted bits as a function of the error probability that her eavesdropping creates.  Upper curve:  Fuchs-Peres-Brandt probe with minimum-error-probability projective measurement.  Lower curve:  Brandt's conclusive entangling probe with POVM.}
\end{figure}

The preceding performance comparison is unfair to \cite{QIP}--\cite{arx2}, in that these papers offer the conclusive entangling probe as a means to make BB84 QKD vulnerable when there is loss between Alice's transmitter and Bob's receiver.  As an extreme version of the scenario contemplated in these works, suppose that:  Alice's transmitter and Bob's receiver are connected by a pure-loss (no excess noise) channel of transmissivity $0 < \eta < 1$; Eve connects Alice's transmitter to the control qubit of her CNOT gate by a lossless channel; and Eve can send photons on to Bob's receiver through another lossless channel.  Then, according to \cite{QIP}--\cite{arx2}, Eve can use a conclusive entangling probe with $\Pr(e = {\rm inc}) = 1 - \eta$, get error-free measurements on all sifted bits for which her outcome is not $T_{\rm inc}$, and route to Bob---through her lossless channel---only the control-qubit output photons for which her measurement was \em not\/\rm\ $T_{\rm inc}$.  For this arrangement, \cite{QIP}--\cite{arx2} claim that Bob would obtain the same photon flux, and Alice and Bob would have no errors on their sifted bits, whether or not Eve's conclusive entangling probe was present.  In addition, according to \cite{QIP}--\cite{arx2}, Eve would obtain perfect information about each of those error-free sifted bits when she used that probe.  However, because the analysis in \cite{QIP}--\cite{arx2} has neglected the $|T_E\rangle$ component of Eve's target-qubit output state, the preceding conclusion is incorrect, as we will now show.   

Eve's R\'{e}nyi information about Bob's error-free sifted bits, in the preceding scenario, satisfies
\begin{eqnarray}
I_R &\equiv&
-\log_2\!\left(\sum_{b=0}^1[\Pr(\,b\mid e \neq {\rm inc}, b=a\,)]^2\right) \nonumber \\
&+&\sum_{e=0}^1\Pr(\,e\mid e \neq {\rm inc}, b=a\,) \nonumber \\ 
&\times& \log_2\!\left(\sum_{b=0}^1 [\Pr(\,b\mid e, e \neq {\rm inc}, b=a\,)]^2\right).
\end{eqnarray}
The probabilities needed to evaluate this expression follow readily from Table~II:
\begin{equation}
\Pr(\,b = j \mid e \neq {\rm inc}, b = a\,) = 1/2,\quad\mbox{for $j=0,1$}
\end{equation}
\begin{equation}
\Pr(\,e = k \mid e \neq {\rm inc}, b = a\,) = 1/2,\quad\mbox{for $k = 0,1$}
\end{equation}
and
\begin{equation}
\Pr(\,b = j\mid e = j, e \neq {\rm inc}, b=a\,) = 1,\quad\mbox{for $j=0,1$},
\end{equation}
from which we get
\begin{equation}
I_R = 1,
\end{equation}
in agreement with \cite{QIP}--\cite{arx2}.  However, this perfect R\'{e}nyi information does \em not\/\rm\ come for free.  In a final application of Table~II we find that the Alice/Bob sifted-bit error probability for this conclusive-probe scenario is
\begin{equation}
\Pr(\,b \neq a \mid e \neq {\rm inc}\,) = 1/3.
\end{equation}
So, this form of the conclusive entangling probe achieves the \em same\/\rm\ error-probability/R\'{e}nyi-information trade-off as the FPB probe with $P_E = 1/3$.  Hence it does \em not\/\rm\ represent a hitherto unexpected vulnerability of BB84 QKD.  

\acknowledgments
This work was supported by the Department of Defense Multidisciplinary University Research Initiative program under Army Research Office grant DAAD-19-00-1-0177.

\end{document}